\documentclass[prl,twocolumn,aps,amssymb,showpacs,superscriptaddress]{revtex4}
\usepackage{graphicx}
\usepackage{color}

\begin{document}

\title{Quasiscarred modes and their branching behavior at an exceptional point}
\preprint{prepared for PRE}

\author{Jung-Wan Ryu}
\address{Max-Planck Institute for Physics of Complex Systems, N\"{o}thnitzer Strasse 38, Dresden, Germany}

\author{Soo-Young Lee}
\email{pmzsyl@phya.snu.ac.kr}
\address{School of Physics and Astronomy, Seoul National University, Seoul 151-742, Korea}

\date{\today}

\begin{abstract}
We study quasiscarring phenomenon and mode branching at an exceptional point (EP) in typically deformed microcavities. It is shown that quasiscarred (QS) modes are dominant in some mode group and their pattern can be understood by short-time ray dynamics near the critical line. As cavity deformation increases, high-Q and low-Q QS modes are branching in an opposite way, at an EP, into two robust mode types showing QS and diamond patterns, respectively. Similar branching behavior can be also found at another EP appearing at a higher deformation. This branching behavior of QS modes has its origin on the fact that an EP is a square-root branch point.
\end{abstract}


\maketitle

Open quantum/wave systems often exhibit interesting phenomena that have no correspondence
in closed systems. Openness effects of the systems are important to understand the physics of them. Quasiscarring phenomenon is an example found in a spiral-shaped optical microcavity \cite{LeeSY04}. Unlike the scarred mode \cite{Heller84,LeeSY05}, the quasiscarred (QS) mode shows amplitude enhancement along a closed path, {\it not a periodic orbit}, due to the openness effects such as Goos-H\"{a}chen (G-H) shift \cite{Goos47,Hentschel02} and Fresnel filtering (FF) effect \cite{Rex02,Tureci02}. Although QS modes have been found only in the spiral-shaped microcavity, the openness effects necessary for quasiscarring are so generic that QS modes should be found in other deformed microcavities.

On the other hand, even for integrable systems, the openness explains the localized mode pattern along periodic orbits \cite{Unter08}. It is also noted that the openness effects responsible for quasiscarring still exist even in the regular regime.  Therefore, it is natural to expect the existence of QS modes in  a slightly deformed microcavity showing almost regular ray dynamics.

One of inherent features of open quantum/wave systems is the non-trivial energy surface near a degenerate point, called exceptional point (EP) \cite{Kato66,Heiss00}, at which two eigenvalues and eigenfunctions coalesce. The interesting physics related to the EP has been studied in various open systems \cite{Dembowski01,Cartarius07,Choi10}, and recently an EP has been found in optical microcavities \cite{LeeSY08,Ryu09,LeeSB09}. The EP is a square-root branch point mathematically, so neighboring two modes can become quite different, as belonging to different branches, on a passage by an EP \cite{LeeSY08}. 

In this paper, we study QS modes and their branching behavior at an EP in typically deformed microcavities, quadrupole and stadium deformations. The QS pattern can be understood by short-time ray dynamics near the critical line, in both chaotic and almost regular regimes. It is also shown that the whispering-gallery modes (WGMs) localized around the critical line are continued into either the QS branch or the diamond branch as the deformation increases. A degenerate point (EP) of high-Q and low-Q QS modes is found to play a crucial role in the mode branching.

\begin{figure}
\includegraphics[width=3.5in]{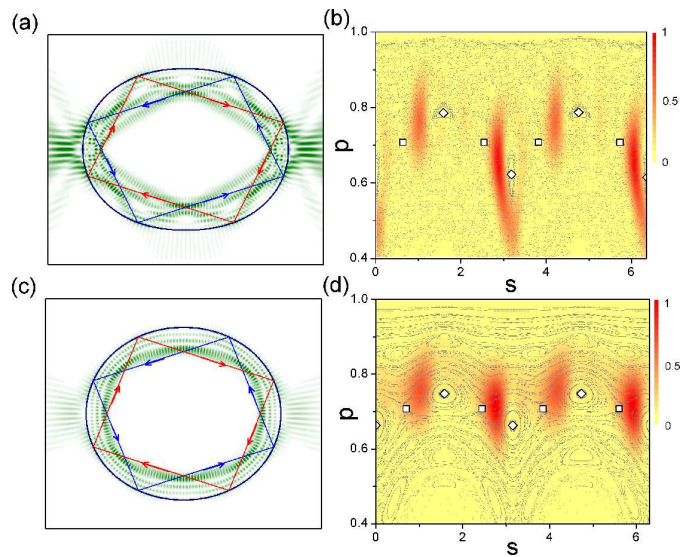}
\caption{(Color online) (a) A QS mode in a chaotic microcavity. ($A_2$ mode at $\varepsilon=0.12$) (b) Husimi function of the QS mode shown in (a). (c) A high-Q QS mode in an almost regular microcavity. ($A_2$ mode at $\varepsilon=0.06$) (d) Husimi function of the QS mode shown in (c). In (b) and (d), the PSOSs are shown in background and the diamonds and rectangles represent diamond and rectangle periodic orbits.
}
\end{figure}

Although the quasiscarring has been studied only in a spiral boundary shape \cite{LeeSY04,LeeJ08,Kim09}, we can expect that QS modes, formed by openness effects, would exist in other deformed microcavities. We thus take a typical deformation, quadrupolar deformation, given as $r(\phi)=R(1+\varepsilon \cos 2 \phi )$ in polar coordinate with $R=1$. We set $n=\sqrt{2}$, $n$ the refractive index, in order to study quadrangle QS modes by noting the maximum openness effects around the critical incident angle $\chi_c=\arcsin (1/n)$ \cite{LeeSY05}.
In this paper, we focus on TM(transverse magnetic) polarization in which wave function and its normal derivative are continuous across cavity boundary.
 Mode solutions are obtained by the boundary element method \cite{Wiersig03}. Figure 1 (a) shows a QS mode found at a deformation $\varepsilon =0.12$. The internal pattern looks somewhat complicated due to the interference between clockwise (red arrows) and counter-clockwise (blue arrows) waves propagating along each quadrangle path. The bouncing positions are clearly seen in the Husimi function for incident waves \cite{Hentschel03} as shown in Fig. 1 (b) where only the counter-clockwise part, $p>0$, is shown for a clear view, $p=\sin \chi$, $\chi$ is incident angle and $s$ is the boundary coordinate. It is evident that the bouncing positions locate between two periodic orbits, rectangle and diamond orbits (see rectangles and diamonds), so this is a QS mode. The Poincar\'{e} surface of section (PSOS) plotted in background shows a chaotic ray dynamics.

We note that the openness effects, such as G-H shift and FF effect, needed in the quasiscarring still exist even when ray dynamics is almost regular. This means that the quasiscarring does not demand the chaotic ray dynamics and QS modes can be found in the small deformation regime with almost regular phase space. This is the case shown in Fig. 1 (c), a QS mode at $\varepsilon=0.06$. The bouncing positions shown in Fig. 1 (d) clearly show that the corresponding quadrangle is not a periodic orbit. One can see almost regular ray dynamics from the PSOS in background. It is emphasized that the QS modes observed cannot be explained as a scarred mode modified by openness effect such as G-H shift and FF effect which give small shifts of a wavelength order and a degree in incident angle \cite{LeeSY05}. Therefore, as in the spiral-shaped cavity there is no periodic orbit associated with the QS mode pattern.

\begin{figure}
 \includegraphics[width=3.5in]{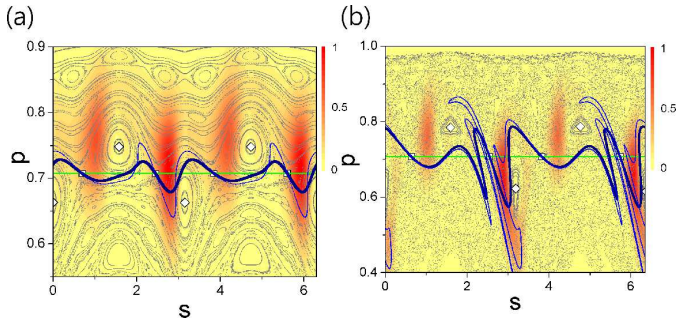}
\caption{(Color online) Short-time evolution of the critical line. The thick and thin lines are the images of the critical line (green straight line) after two and four bounces, respectively. The good agreement with Husimi function below the critical line has an implication for the QS-mode formation. (a) $\varepsilon=0.06$. (b) $\varepsilon=0.12$.
}
\end{figure}

 The bouncing points of counter-clockwise waves in QS modes of Figs. 1 (b) and (d) locate on the {\it left} side of those of the diamond orbit. Why not on the right side? In order to understand this point, we have to take short-time ray dynamics into account. It is emphasized that even in a chaotic case rays being above the critical line come down below the line and escape following some dynamical passages associated with unstable-manifold structure \cite{LeeSY05,Schwefel04}.
Let us consider a ray ensemble evenly distributed above the critical line and see how the rays evolve in a short-time period, especially near the critical line where QS modes are localized. It can be illustrated by evolution of the critical line as shown in Fig. 2.
The green and yellow lines denote the images of the critical line after two and four bounces, respectively. It is clear that the passages to escape locate on the {\it left} side of the diamond orbit, and they are in a good agreement with the bouncing points of the QS quadrangles. This implies that the pattern of the QS modes is the consequence of a natural constructive interference, under openness effects, of the waves coming down from the upper part of the critical line.

\begin{figure}
 \includegraphics[width=3.5in]{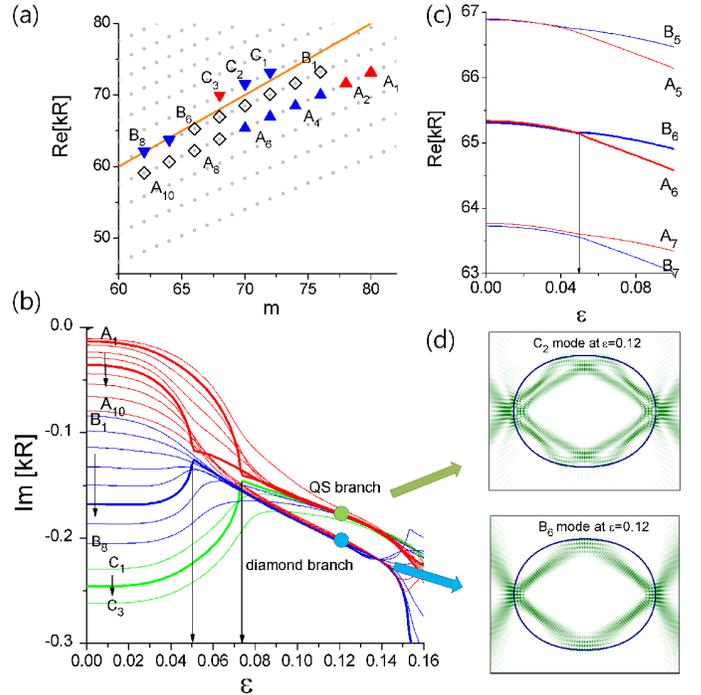}
\caption{(Color online) (a) All existing modes (gray dots) in a circular case and the modes of even-even symmetry class, $(A_1, \cdots, A_{10}, B_1, \cdots, B_8, C_1, \cdots, C_3)$. They are taken as initial modes for the evolution study of (b). Diamonds and triangles imply the mode patterns at a larger deformation, diamond and QS mode patterns, respectively. (b) Evolution of Im[$kR$] of the chosen initial modes as increasing $\varepsilon$. Existence of an EP at $\varepsilon \simeq 0.05$ is seen as a singular-like behavior of $A_6$ and $B_6$, and another EP around $\varepsilon \simeq 0.073$ (see bold lines and arrows). (c) Transition from crossing of $(A_5,B_5)$ and $(A_6,B_6)$ to avoided crossing of $(A_7,B_7)$ of Re[$kR$] at the EP. (d) Mode patterns of a QS mode and a diamond mode, corresponding to green and blue dots in (b).
}
\end{figure}

In fact, the appearance of QS modes shown in Fig. 1 is not exceptional, but rather common in typically deformed microcavities. In order to see their common existence, we investigate evolution of the WGM modes with incident angles $\chi$ similar to the critical angle $\chi_c$ for total internal reflection in the range of $52 \le m \le 80$, $m$ is the angular quantum number ($m$ is even in our case of even-even symmetry class). In Fig. 3 (a) small dots are all WGMs existing in the range shown and each mode has two quantum number $(m,l)$, $l$ is the radial quantum number. The modes of the lowest Re[$kR$], $k$ the wavenumber, at each $m$ belong to $l=1$ mode group, and the next-lowest ones to $l=2$ mode group and so on. 
The orange line denotes Re[$kR$]=$m$ corresponding to the critical incident angle $\chi_c=\arcsin (1/n)$. For convenience, we mark the modes of $l=4$ as $A_1,\cdots,A_{10}$, the $l=5$ modes as $B_1, \cdots, B_8$, and the $l=6$ modes as $C_1,\cdots,C_3$ as indicated by diamonds and triangles in Fig. 3 (a).

Evolutions of these modes with $\varepsilon$ are well characterized by the plot of Im[$kR$], related to decay rate, as shown in Fig. 3 (b). The red lines denote evolutions of $A_i$ modes, and the blue and green lines do those of $B_i$ and $C_i$ modes, respectively. From a low deformation about $\varepsilon \simeq 0.02$ almost modes begin to exhibit QS character, and we will call them high-Q QS modes if their incident angle is greater than $\chi_c$ and low-Q QS modes otherwise, some of which are indicated by upward-pointing and downward-pointing triangles, respectively, in Fig. 3 (a). The Im[$kR$] values of these  QS modes converge and are connected to two branches after $\varepsilon \sim 0.08$, the modes in the upper branch exhibit a QS mode pattern and others in the lower branch do a diamond mode pattern. Figure 3 (d) shows these patterns corresponding to $C_2$ mode and $B_6$ mode at $\varepsilon =0.12$ (green dot and blue dot in Fig. 3 (b)).

Another interesting feature is the existence of two exceptional points (EPs) near $\varepsilon \simeq 0.05$ and $\varepsilon \simeq 0.073$. The EP is a degenerate point in a non-Hermitian system like an open quantum system, at which two eigenvalues and eigenfunctions coalesce \cite{Heiss00}. Mathematically, the EP is a square-root branch point in a parameter space \cite{LeeSY08}.
 To find an EP, we need two parameters, i.e., its codimension is two. In our case we have quasi-two dimensional parameter space, one is the deformation $\varepsilon$ and the other is a discrete internal parameter $m$ \cite{LeeSB09,LeeSB09R}. Due to the discreteness of $m$, we cannot find the exact EP position, but can recognize the vicinity of EP from the singular-like behavior of the evolutions. The evolution lines of $A_6$ and $B_6$ modes show abrupt changes at $\varepsilon \simeq 0.05$, and those of $A_2$ and $C_2$ modes do at $\varepsilon \simeq 0.073$ (see bold lines in Fig. 3 (b)). These EPs can be also confirmed by the transition behavior from crossing to avoided crossing in Re[$kR$] plot, as shown in Fig. 3 (c), as decreasing $m$. We also see that $kR$ values of $A_6$ and $B_6$ are very close at $\varepsilon \simeq 0.05$. The same behavior appears near $\varepsilon \simeq 0.073$ in the $A_2$ and $C_2$ evolutions. In fact, by introducing another continuous parameter like the refractive index $n$, we can find nearby EPs, in our case, EPs locate at $(n_{EP},\varepsilon_{EP}) \simeq (1.4164,0.050)$ and $(1.4160,0.0730)$.

Due to the nontrivial topology of energy surfaces near an EP, neighboring two modes showing almost identical mode properties can be completely different modes as a parameter passes by an EP \cite{Heiss00,LeeSY08}. This branching property is observed in Fig. 3 (b). For example, the EP existing at $\varepsilon_{EP} \simeq 0.05$ divides similar modes, $(A_6,A_7)$ or $(B_6,B_7)$, at $\varepsilon < \varepsilon_{EP}$ into different modes, a QS mode $(A_6,B_7)$ and a diamond mode $(A_7,B_6)$, at $\varepsilon > \varepsilon_{EP}$. As a result, while $(A_3, \cdots , A_6,B_7,B_8)$ modes evolve into the QS branch, $(A_7, \cdots, A_{10},B_1, \cdots, B_6)$ modes do into the diamond branch, as marked in Fig. 3 (a) by triangles and diamonds. Similar behavior is expected to happen by the other EP at $\varepsilon_{EP} \simeq 0.73$, while in Fig. 3 (b) the modes passing by the EP look approaching the same QS mode branch. Actually, looking into details, there are two kinds of QS modes at $\varepsilon > \varepsilon_{EP}(\simeq 0.73)$. They have a small difference in bouncing positions and Q factor, so we say QS-I modes and QS-II modes for convenience. The QS-I mode at $\varepsilon = 0.12$ has the pattern shown in Fig. 1(a), and the other QS-II mode has slightly different pattern shown in the upper panel of Fig. 3 (d). Therefore, $(A_1,A_2, C_3)$ and $(A_3,\cdots,A_6,C_1,C_2)$ become the QS-I  and the QS-II modes, respectively, as marked by red and blue colors in Fig. 3 (a).

\begin{figure}
\hspace{-0.5cm} \includegraphics[width=3.6in]{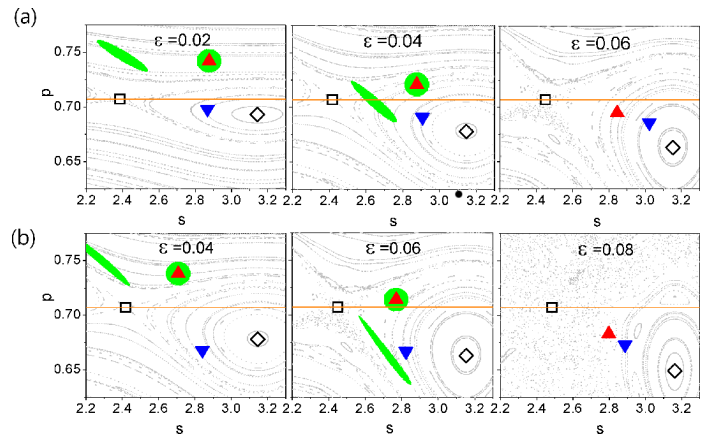}
\caption{(Color online) Change of peak positions of Husimi functions for two interacting QS modes, the high-Q QS mode (red upward-pointing triangle) and the low-Q QS mode (blue downward-pointing triangle) before and after $\varepsilon=\varepsilon_{EP}$. The ray return image of the high-Q QS mode point marked by green circle is shown. (a) $\varepsilon_{EP} \simeq 0.05$ case of $A_6$ and $B_6$ modes. (b) $\varepsilon_{EP}\simeq 0.073$ case of $A_2$ and $C_2$ modes. The critical lines are denoted by orange lines.
}
\end{figure}

The interaction between a high-Q and a low-Q QS modes can be modeled by a two-by-two interaction Hamiltonian matrix with diagonal elements $E_{\{H,L\}}=\omega_{\{H,L\}}-i\gamma_{\{H,L\}}/2$, $\omega$ is the frequency and $\gamma$ the decay rate, and the off-diagonal element ${\cal C}_{HL}$ representing coupling strength between them \cite{LeeSB09}. The condition for an EP is then given by ${\cal C}_{HL}=|\gamma_H-\gamma_L|/4$ when $\omega_H=\omega_L$. In our case, all elements change depending on $\varepsilon$. For a small deformation, $|\gamma_H-\gamma_L|$ is much greater than 4${\cal C}_{HL}$, similar to an integrable case. As increasing deformation, not only $|\gamma_H-\gamma_L|$ decreases (see Fig. 3 (b)) but also the ${\cal C}_{HL}$ increases due to introduction of nonintegrability of ray dynamics. As a result,  at some deformation the condition ${\cal C}_{HL}=|\gamma_H-\gamma_L|/4$ would be hold, and the point would be an EP if $\omega_H=\omega_L$ at the deformation. Therefore, the EP can exist roughly when the coupling ${\cal C}_{HL}$ increases drastically.

The change of ${\cal C}_{HL}$ can be qualitatively estimated from the ray return image of the high-Q QS pattern. For example, consider $A_6$ mode at $\varepsilon = 0.02$ whose a peak position of Husimi function is the red upward-pointing triangle in Fig. 4 (a). The rays at the position (see the green circle) return, after four bounces, to the left side of the initial position as indicated by the green image, implying a small coupling ${\cal C}_{HL}$ to the low-Q mode ($B_6$ mode) existing below (see the blue downward-pointing triangle). At a higher deformation $\varepsilon=0.04$, the return image locates below the initial point due to the widen nonlinear resonance. This indicates more larger coupling with the low-Q QS mode since wave functions of $A_6$ and $B_6$ modes are expected to show a larger overlap. This increase of ${\cal C}_{HL}$ is consistent with the existence of an EP at $\varepsilon\simeq 0.05$. After the EP, e.g., at $\varepsilon=0.06$, the peak points, initially showing a vertical difference (a large Q difference), distribute with a horizontal difference (a small Q difference), the high-Q QS mode becoming a QS-II mode and the low-Q QS mode approaching a diamond mode. The peak variations and the return image of $A_2$ and $C_2$ modes exhibit the same behavior, as shown in Fig. 4 (b), except for the different final modes, QS-I and QS-II modes.

\begin{figure}
\includegraphics[width=3.5in]{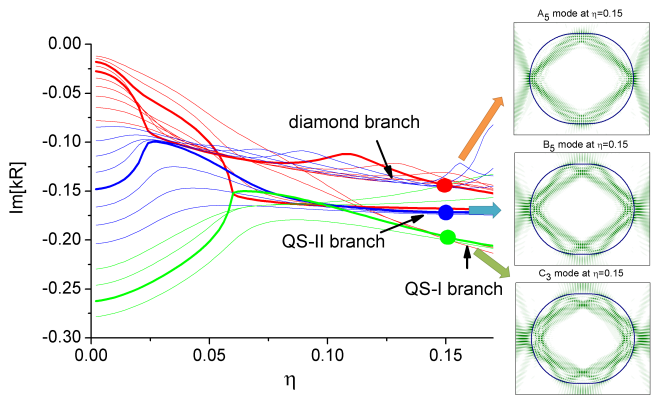}
\caption{(Color online) Mode evolutions in the stadium deformation. All modes can be classified into three modes, a diamond scar mode and two QS modes whose patterns are shown at the right panels.
}
\end{figure}

The branching role of the EPs for QS modes are quite generic. As shown in Fig. 5, a similar evolution behavior of Im[$kR$] can be found in the stadium deformation showing a fully chaotic ray dynamics. The deformation parameter $\eta$ is defined by $\eta=a/2R$, $a$ is the length of the parallel boundary and $R$ is the radius of circular boundary. In this stadium case, $A_5$ and $B_5$ mode evolve through the vicinity of an EP located around $\eta=0.025$ associated with two branches, a diamond scar branch and QS-II branch, as shown in Fig. 5.  $A_3$ and $C_3$ modes also pass by another EP near $\eta=0.06$ related to two QS branches, QS-II and QS-I branches. The modes patterns of the three branches are shown in Fig. 5. This branching behavior by two EPs is similar to the quadrupole case, i.e., $(A_1,A_2,C_3,C_4)$, $(A_3,A_4,B_5,\cdots,B_8,C_1,C_2)$, and $(A_5,\cdots,A_{10}, B_1,\cdots,B_4)$ modes arrive at the QS-I, the QS-II, and the diamond branches, respectively. Note that the modes in the diamond branch have higher Q values, unlike the quadrupole case, due to the different boundary geometry, and show small wiggles implying interactions with other high-Q modes. This similar branching behavior of the stadium case means that the interaction between QS modes is not directly relevant to the PSOS structure. Instead, the combination of openness and short-time ray dynamics is more responsible for the interaction. Other branching behavior by an EP is expected for other modes since it is
based on the singular feature of the EP.

In conclusion, we have shown that appearance of QS modes is common in a typically deformed microcavity such as a quadrupole and a stadium microcavities. At a small deformation, they can be divided into high-Q and low-Q QS modes, and both QS modes evolve into three branches, a diamond branch and two QS (QS-I,QS-II) branches. This branching behavior has been understood by the existence of two EPs, i.e., the inherent property of the square-root branch points.
The mode pattern of a QS mode can be explained by a combination of openness and short-time ray dynamics near the critical line, rather than its PSOS structure, due to its leaky nature.

This work was supported by NRF grant (2010-0008669).

\end{document}